\documentstyle[prl,aps,twocolumn,epsfig]{revtex}
\begin{document}
\twocolumn[\hsize\textwidth\columnwidth\hsize\csname @twocolumnfalse\endcsname

\title{Phase diagram of the relaxor ferroelectric $(1-x)$Pb(Zn$_{1/3}$Nb$_{2/3}$)O$_{3}$-$x$PbTiO$_{3}$}
\author{D. La-Orauttapong$^{1}$, B. Noheda$^{2}$, Z.-G. Ye$^{3}$, P.M. Gehring$^{4}$, J. Toulouse$^{1}$, D.E. Cox$^{2}$, and G. Shirane$^{2}$}
\address{$^{1}$Department of Physics, Lehigh University, Bethlehem, PA 18015, USA}
\address{$^{2}$Brookhaven National Laboratory, Upton, NY 11973, USA}
\address{$^{3}$Department of Chemistry, Simon Fraser University, Burnaby, BC V5A 1S6, Canada}
\address{$^{4}$NIST Center for Neutron Research, NIST, Gaithersburg, MD 20899, USA}
\maketitle

\begin{abstract}
Recently, a new orthorhombic phase has been discovered in the ferroelectric 
system $(1-x)$Pb(Zn$_{1/3}$Nb$_{2/3}$)O$_{3}$-$x$PbTiO$_{3}$ (PZN-$x$PT) for $x$= 9\%, 
and for $x$= 8\% after the application of an electric field. 
In the present work, synchrotron x-ray measurements have been extended to higher 
concentrations, 10\% $\leq x \leq $15\%. The orthorhombic phase was observed 
for $x$= 10\%, but, surprisingly, for $x\geq $11\% only a tetragonal phase was found down to 20 K. The orthorhombic phase thus exists only in a narrow concentration range with near 
vertical phase boundaries on both sides. This orthorhombic symmetry 
(M$_C$-type) is in contrast to the monoclinic M$_A$-type symmetry recently identified at 
low temperatures in the Pb(Zr$_{1-x}$Ti$_{x}$)O$_{3}$ (PZT) system over 
a triangle-shaped region of the phase diagram in the range 
$x$= 0.46-0.52. To further characterize this relaxor-type system neutron 
inelastic scattering measurements have also been performed 
on a crystal of PZN-$x$PT with $x$= 15\%. The anomalous soft-phonon behaviour (``waterfall'' effect) previously observed 
for $x$= 0\% and 8\% is clearly observed for the 15\% crystal, which indicates that 
the presence of polar nanoregions extends to large values of $x$. 
\end{abstract}

%\vskip5pc
\vskip1pc]

\narrowtext

\section{Introduction}
Recently, a number of studies have attempted to understand the origin of the
very large piezoelectric coefficients measured in perovskite oxides such as
Pb(Zr$_{1-x}$Ti$_{x}$)O$_{3}$ (PZT) and $(1-x)$Pb(Zn$_{1/3}$Nb$_{2/3}$)O$_{3}
$-$x$PbTiO$_{3}$ (PZN-$x$PT) near the morphotropic phase boundary (MPB).
The MPB is an almost vertical phase boundary that separates the rhombohedral
(R) and the tetragonal (T) regions of the phase diagram of these systems
(temperature vs $x$). In a study of PZN-$8\%$PT, Park and Shrout found the
piezoelectric coefficient $d_{33}$ to exceed 2500 pC/N and strain levels
reaching 1.7$\%$ induced by a field applied along [001] \cite
{Park_et_al_1804_1997}. These ultrahigh values are an order-of-magnitude
greater than those previously attainable in conventional piezoelectric and
electrostrictive ceramics including PZT, currently the material of choice
for high performance actuators. Several experimental and theoretical
studies now indicate that these very high values are related to the presence
of a particular phase.

X-ray investigations by Noheda et al.\cite{Noheda_et_al_PRB_14103_2001} and
Cox et al.\cite{Cox_et_al} have shown that, in addition to the known
rhombohedral and tetragonal phases, a sliver of a new phase exists in the
phase diagram near the MPB as shown in Fig.1 for PZT \cite
{Noheda_et_al_PRB_14103_2001} and PZN-$x$PT\cite{Cox_et_al}, respectively. 
In the PZT system, the newly identified lower-symmetry phase is of
monoclinic M$_{A}$ type (space group Cm) for $0.46 \leq x \leq 0.52$ \cite
{Noheda_et_al_PRB_14103_2001}, while in PZN-$x$PT an orthorhombic (O)
phase (space group Bmm2) has been irreversibly induced by a field in an
8\%PT crystal \cite{Noheda_et_al_3891_2001,Viehland} and has also been
observed in a polycrystalline sample prepared from a previously-poled
crystal of 9\%PT\cite{Cox_et_al}. This O phase is the limiting case of a
monoclinic M$_{C}$ type phase (space group Pm) when the lattice parameters a
and c become equal. A true M$_{C}$ phase (with a $\neq c)$ is observed in
PZN-8\%PT during the application of an electric field \cite
{Noheda_et_al_3891_2001}. Very recently, Uesu et al. \cite
{Uesu_et_al._Cond-mat/00106552} have also observed a true M$_{C}$ phase in
9\%PT following an examination of several 

\begin{figure}[tb]
\epsfig{width=0.8\linewidth,figure=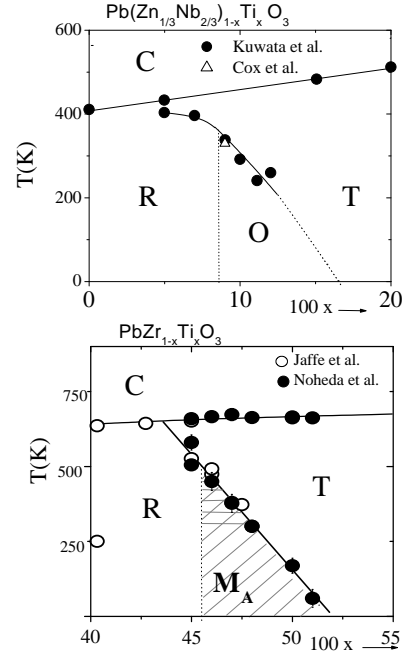}
\caption{Phase diagrams for PZT (bottom) and PZN-$x$PT (top) in the vicinity
of their respective MPB's, as shown in Ref. \protect\onlinecite{Cox_et_al}}
\label{fig1}
\end{figure}

\noindent unpoled (as-grown) as well as poled crystals with nominally the same 9\% concentration. Most of these were orthorhombic O but one of the unpoled crystals showed a definite monoclinic M$_{C}$ distortion. \ This raises the very interesting possibility that, for slightly higher concentrations (e.g. 10-12\%PT), the ground state is actually M$_{C}$, so that the orthorhombic 9\%PT phase might logically be regarded as the 
end-member of the M$_{C}$ region. However, as described below, the 10\%PT phase has been found to be unequivocally orthorhombic ($a=c$), while unexpectedly a tetragonal phase is realized at 11\%PT, yielding a narrow ``chimney-like'' shape for the intermediate 
orthorhombic region.

A low-symmetry phase discovered in both PZT and PZN-$x$PT in the vicinity
of the MPB appears to be a common feature of the highly-piezoelectric
perovskite systems. \ Recently, Vanderbilt and Cohen \cite
{Vanderbilt_et_al_PRB_94108_2001}, hereafter referred to as VC, have
provided a natural explanation for these recently-discovered new phases in
both PZT and PZN-$x$PT systems by extending the Devonshire theory to
eighth-order. This study yields a new phase diagram for ferroelectric
perovskites that includes three different types of monoclinic phases M$_{A}$%
, M$_{B}$, and M$_{C}$ (named after this work).

The effect of an applied electric field on the polarization has also been
studied theoretically and experimentally. A polarization rotation
mechanism has been proposed by Fu and Cohen\cite{Fu_et_al_281_2000} to
explain the ultrahigh electromechanical response found in PZN-$x$PT using
BaTiO$_{3}$ as a model.\ According to their model, the application of an
electric field along the [001] direction in rhombohedral PZN-$x$PT induces a
rotation of the polarization vector in a (110) plane, from the rhombohedral
to the tetragonal axis (i.e., R-M$_{A}$-T).\ However, an experimental x-ray
study by Noheda et al. on as-grown PZN-$x$PT crystals has suggested that, as
the applied electric field is increased, the polarization vector first
follows this R-M$_{A}$-T path but then abruptly jumps to a new path, in a
plane containing the orthorhombic and tetragonal polar axes (i.e., R-M$_{A}$%
-M$_{C}$-T). \cite{Noheda_et_al_3891_2001} \ As the field is decreased the
polarization rotates from the tetragonal [001] to the orthorhombic [101]
polar directions, via the M$_{C}$ phase, and the initial rhombohedral state
is not recovered upon removal of the field.\ Recently this irreversible R-M$%
_{A}$-M$_{C}$-T polarization path has been confirmed by neutron diffraction. 
\cite{Ohwada_et_al}\ \ First principle calculations by Bellaiche et al. for
the rhombohedral PZT system under a [001] field also predict the R-M$_{A}$-M$%
_{C}$-T transformation, although in this case the transformation is found to
be reversible. \cite{Bellaiche_et_al}

With ample experimental and theoretical evidence \cite
{Guo_et_al_5423_2000,Bellaiche_et_al_5427_2000} for a link between the very
high values of the piezoelectric and electrostrictive coefficients and the
presence of a new phase, it has become essential to determine its extent 
in the phase diagram of PZN-$x$PT. \ As mentioned above,
PZN-8\%PT shows the expected rhombohedral symmetry and becomes orthorhombic
only after a high field has been applied, whereas PZN-9\%PT clearly shows
the orthorhombic symmetry (M$_{C}$ with $a=c$ ) even with no external 
field.\cite{Cox_et_al,Uesu_et_al._Cond-mat/00106552} \ In the present
paper, we have studied higher PT concentrations ($10\%\leq x\leq 15\%$),
using high resolution synchrotron x-ray powder diffraction. \ For further
characterization of these relaxor systems, neutron inelastic scattering
measurements have also been performed on the 15\%PT crystal. \ The specific
goal of the inelastic measurements was to investigate the possible
existence, at higher PT concentrations, of the ``waterfall '' shape
previously observed in the soft optic mode phonon branch at lower PT
concentrations.\cite{Gehring_et_al_5216_2000,Gehring_et_al_224109_2001}
\bigskip \bigskip

\section{Experiment}

Single crystals of (1-$x$)Pb(Zn$_{1/3}$Nb$_{2/3}$)O$_{3}$ - $x$PbTiO$_{3}$
solid solution system with $x$ = 10, 11, 12 and 15\% were grown by a top
seeded solution growth (TSSG) technique using a PbO flux with an optimum
flux ratio of 50 wt\%.\cite{Chen_et_al_inpress} \ Platelets of about 1 mm
thick and 7 to 15 mm$^{2}$ in area were cut parallel to the reference (001)$_{%
\text{cub}}$ plane and polished with fine diamond paste (down to $1 \mu $m). 
The (001)$_{cub}$ faces were covered with sputtered gold layers and
Au-wires were attached by means of Ag-paste. \ The poling was
performed under a field of 20 kV/cm that was applied along [001]$_{cub}$ 
at 210${^{\circ }}$C (above T$_{C}$) and maintained while cooling
down to room temperature. The samples were then short-circuited for 30 min.
before the electrodes were removed. \ For neutron inelastic scattering
studies, a bigger PZN-15\%PT crystal of 2.17 g in weight and 0.27 cc in
volume was cut from an as-grown crystal boule with natural (100)$_{cub}$ faces 
as a reference orientation.

Two types of high-resolution synchrotron x-ray powder diffraction
measurements were carried out on beam line X7A at the Brookhaven National
Synchrotron Light Source (NSLS) with x-rays from a Si(111) double-crystal
monochromator. \ For the first set of measurements, an incident beam of 
wavelength $\sim $0.7 \AA\ was used and a linear position-sensitive detector 
was placed in the diffracted beam path. This configuration gives greatly enhanced
counting rates and allows accurate data to be collected from very
narrow-diameter capillary samples. The use of capillary samples helped
eliminate systematic errors due to preferred orientation or texture effects.
With this configuration the resulting instrumental resolution was $\sim $0.03$%
{^{\circ }}$ on the $2\theta $ scale.\ For the second set, incident
beams of wavelengths $\sim $0.69 \AA\ and $\sim $0.98 \AA\ were used with
a flat Ge (220) crystal-analyzer and scintillation detector.\ With this type
of diffraction geometry, it is not always possible to eliminate preferred
orientation and texture effects, but the peak positions, on which the
present results are based, are not affected. The resulting instrumental
resolution is better than 0.01${^{\circ }}$ on a $2\theta $ scale, an
order of magnitude better than that of a laboratory instrument.

The powder samples were prepared by chopping out a small fragment from the
poled crystals, crushing them and loading them in a 0.2mm diameter glass
capillary, as explained in Ref.\protect\onlinecite{Cox_et_al} A
closed-cycle cryostat was used for the temperature dependence measurements.
The data sets were collected from 15-500 K and the sample was rocked $\sim
2-3$ degrees during the data collection, in order to achieve powder averaging.

A subsequent neutron inelastic scattering study was performed on the
PZN-15\%PT single crystal using both the BT2 and BT9 triple-axis
spectrometers at the NIST Center for Neutron Research (NCNR). The (002)
reflections of highly-oriented pyrolytic graphite (HOPG) crystals were used
for both monochromator and analyzer. An HOPG transmission filter was used
to eliminate 

\begin{figure}[tb]
\epsfig{width=0.85\linewidth,figure=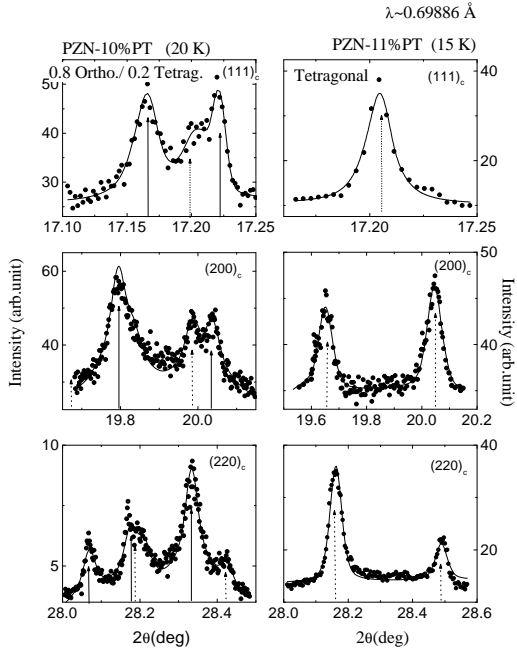}
\caption{Diffraction patterns of pseudo-cubic (111), (200) and (220) for
PZN- $x$PT for $x$ = 10\% (left) at 20 K and $x$ =11\%
(right) at 15 K, showing the diffraction spectra for the orthorhombic and
tetragonal phases.}
\label{fig2}
\end{figure}

\noindent higher-order neutron wavelengths. \ These measurements were made in the 
phonon creation mode with a fixed final energy of 14.7 meV ($\lambda _{f}$=2.36\AA ) 
while varying the incident neutron energy $E_{i}$. \ The horizontal beam collimation 
used was $40{^{\prime }}-40{^{\prime }}-S-40{^{\prime }}$-open and $40{^{\prime }}-
20{^{\prime }}-S-20{^{\prime }}-80{^{\prime }}$. \ The crystal was mounted on a boron 
nitrate cylinder held in a goniometer, and oriented with its [001] axis vertical, 
thereby giving access to the (HK0) scattering zone. \ It was then loaded into a vacuum 
furnace capable of reaching temperatures up to 670K. \ Data were collected in the
temperature range 200-650K using two types of scans. \ First, constant
energy scans were performed by keeping the energy transfer $\hbar \omega $=$%
E_{i}-E_{f}$ fixed while varying the momentum transfer $\overrightarrow{Q}$.
\ Second, constant-$\overrightarrow{Q}$ scans were performed by holding the
momentum transfer $\overrightarrow{Q}$=$\overrightarrow{k}_{i}$-$%
\overrightarrow{k}_{f}$$(k=2\pi /\lambda $) fixed and varying the energy
transfer $\Delta E$.

%====================================Results and discussion==================%

\section{Phase diagram for PZN-{\protect\scriptsize $x$}PT}

Based on the diffraction data reported in the previous study\cite{Cox_et_al}
on a polycrystalline sample of PZN-$x$PT with $x$=9\%, a modification of the
PZN-$x$PT phase diagram 

\begin{figure}[tb]
\epsfig{width=0.95\linewidth,figure=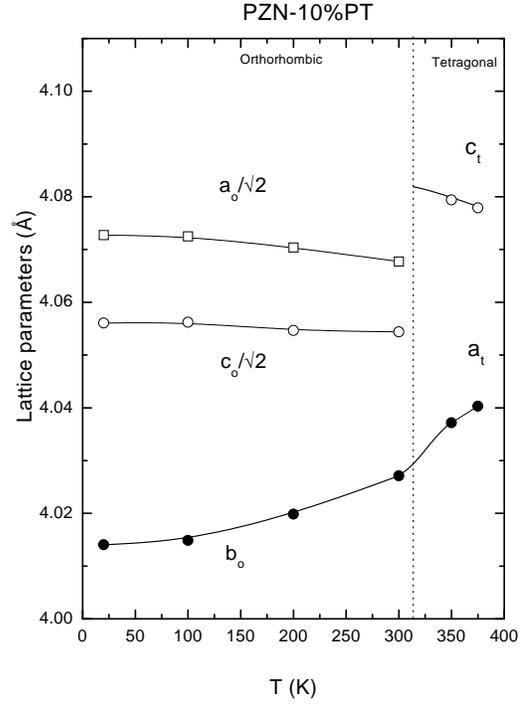}
\caption{Temperature dependence of the lattice parameters for PZN-$x$PT ($x$%
= 10\%) from 15 K to 375 K, for the orthorhombic ($a_{o}$, $b_{o}$ and $c_{o}$) 
and tetragonal ($a_{t}$ and $c_{t}$) phases.}
\label{fig3}
\end{figure}

\noindent has been proposed which includes the new orthorhombic phase (O) around 
its MPB shown in Fig.1. \ However, the proposed stability region for the new O 
phase is limited so far to only one composition.\ In the present work, we have 
determined the full extent of the O phase by studying
higher concentrations, $x$=10, 11, 12, and 15\%. The results obtained
give a comprehensive picture of the new orthorhombic phase in the PZN-$x$PT phase
diagram.

As will be seen later, one of the key features of this study is the rather unexpected 
finding that compositions with $x\geq $11\%  retain tetragonal symmetry down to 20 K. This can be seen in Fig. 2, in which selected regions of the diffraction profiles are plotted for $x$ = 10\% (left) and $x$ = 11\% (right) at 20 and 15 K, respectively, showing very distinctive features for these two adjacent compositions. \ From the
splitting and relative intensities of these profiles the symmetry of the
samples can be reliably established. \ The $x$= 10\% peak profiles are very
well-resolved, and show the same general features previously described for the orthorhombic phase in $x$ = 9\%\cite{Cox_et_al}, but with a small amount of residual tetragonal phase. \ The pseudocubic (111)$_{c}$ reflection consists of two peaks corresponding to the (012) and (210) orthorhombic reflections and a third central peak corresponding to the residual tetragonal (111) singlet. \ Three peaks are observed around the pseudo-cubic (200)$_{c}$ reflection corresponding to the orthorhombic (202)-(020) doublet and the tetragonal (200) reflection. \ The pseudocubic (220)$_{c}$ reflection consists of a triplet, the orthorhombic (004)-(400)-(222), plus the (202)-(220) tetragonal doublet.\ From the intensity ratios one can estimate the fraction of tetragonal 

\begin{figure}[tb]
\epsfig{width=0.95\linewidth,figure=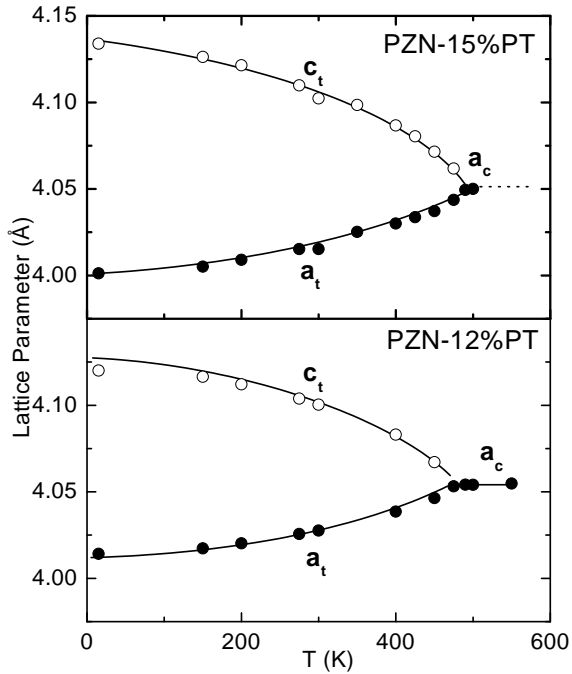}
\caption{Lattice parameters vs. temperature for PZN-$x$PT for $x$ = 15\%
(top) and 12\% (bottom) over the whole temperature range from 15 K to 550 K, 
showing the tetragonal ($a_{t}$ and $c_{t}$) to cubic ($a_{c}$) phase transition.}
\label{fig4}
\end{figure}

\noindent phase to be about 20\% of the sample volume. This fraction is essentially constant between 20-300 K. For the $x$= 11\% composition, the peak profiles are clearly those of a pure tetragonal phase with a single (111) reflection and two doublets,
(002)-(200) and (202)-(220). \ The fits of these profiles show that for $x$ = 10\% \ the orthorhombic unit cell has $a_{o}=5.760$ \AA , $b_{o}=4.014$ \AA, $c_{_{0}}=5.736$ \AA\ and the second tetragonal phase has $a_{t}=$ $4.024$ \AA , $c_{t}$=$4.091$ \AA\ . \ For $x$=11\% the tetragonal unit cell parameters are $a_{t}=4.018$ \AA , $c_{t}=4.113$ \AA .

The evolution of the lattice parameters with temperature for $x$=10\%
reveals an orthorhombic-to-tetragonal phase transition, as shown in Fig.3.\
At increasing temperatures, the orthorhombic lattice parameter $b_{o}$
steadily increases while the $c_{o}/a_{o}$ ratio decreases only slightly.\
At temperatures between 325 and 375 K, the symmetry is found to be
tetragonal with lattice parameters $a_{t}$ and $c_{t}$. The sudden
change in the $a_{o}$ and $c_{o}$ between 300 and 350 K suggests a first-order character to the orthorhombic-tetragonal phase transition. The residual tetragonal phase
which is present between 20-300 K can probably be attributed to small long-range compositional fluctuations in the as-grown crystals. \ The overall behaviour is identical to that previously reported for $x$ = 9\% \cite{Cox_et_al}. \ Similar
temperature-dependence studies have 

\begin{figure}[tb]
\epsfig{width=0.95\linewidth,figure=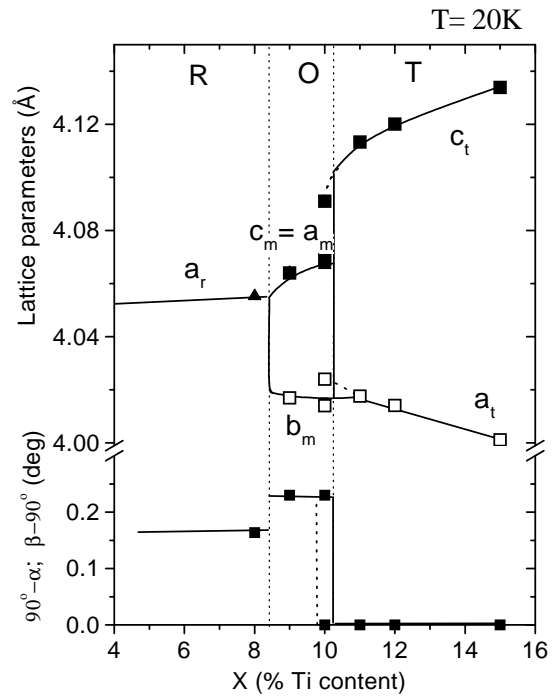}
\caption{Lattice parameters vs. Ti concentration at 20 K for the rhombohedral ($%
a_{r}$ and $\protect\alpha$), monoclinic ($a_{m}$,$b_{m}$, $c_{m}$ and $%
\protect\beta $) and tetragonal ($a_{t}$ and $c_{t}$) phases from the
results of this work, including data from Refs. \protect\onlinecite{Cox_et_al,Noheda_et_al_3891_2001}.}
\label{fig5}
\end{figure}

\noindent been conducted for higher
concentrations and the results are shown in Fig. 4. \ The lattice parameters
as a function of temperature reveal a cubic-tetragonal phase transition for $%
x$ = 15\% (top) and for $x$=12\% (bottom). \ At low temperature, the system
is purely tetragonal. \ With increasing temperature, the strain ratio $%
c_{t}/a_{t}$ decreases, while approaching the transition to the cubic phase at $%
\simeq 490$ K and $\simeq 475$ K, for $x$= 15 and 12\% respectively. 

The lattice parameters as a function of composition at 20 K are presented in
Fig.5, which includes the concentrations studied in this work, as well as
the previous data reported in Refs.\onlinecite{Cox_et_al,Noheda_et_al_3891_2001} for $x$= 8\% and 9\% . \ This figure shows the structural evolution from the rhombohedral to the 
tetragonal phase via the orthorhombic phase. \ For $x\leq 8\%$ the 
crystals exhibit the expected rhombohedral symmetry. \ For $x\geq 11\%$, 
the tetragonal symmetry is found with the strain ratio $c_{t}/a_{t}$ 
increasing from 1.0237 at $x$=11$\%$ to 1.0331 at $x$= 15$\%$. \ The smooth variation in the tetragonal a and c lattice parameters with composition is consistent with fairly precise compositional control during the crystal growth process. \ 
However, between $9\%\leq x\leq 10\%$, the symmetry is found to be orthorhombic.

\section{Soft phonon anomalies in PZN-{\protect\scriptsize $x$}PT}

A series of neutron inelastic measurements were subsequently performed on
the PZN-15\%PT single crystal to study the so-called ``waterfall'' feature,
first observed in PZN-8\%PT\cite{Gehring_et_al_5216_2000}, and later in 
PZN\cite{Gehring_et_al_224109_2001}, at higher concentrations of PbTiO$_3$. \
Specifically, we wished to determine whether or not this waterfall feature persisted 
beyond the MPB. \ The waterfall is an anomaly of the lowest-frequency
transverse optic (TO) phonon branch that is correlated with the
condensation, at $T=T_d$, of local regions of randomly-oriented polarization,
also known as polar nanoregions (PNR), which disrupt the propagation of
long-wavelength TO polar modes. \ The existence of $T_d$, which can be
hundreds of degrees higher than $T_c$, was first reported by Burns and Dacol
for a variety of systems including both PZN and PMN.\cite
{Burn_et_al_853_1983} \ Because the PNR are polar, they naturally couple to
the polar TO phonon mode, which in turn can serve as a microscopic probe of
the PNR. \ Their presence is manifested by a drastic broadening of the TO
phonon energy linewidth below a specific wavevector, and gives rise to a
steep ridge of scattering when plotted as a function of wavevector $q$ as
the long-wavelength phonon cross section becomes distributed in energy. \
The search for the waterfall feature should thus yield important information
about the possible existence of the PNR beyond the MPB.

Examples of two constant-$E$ scans for $\hbar\omega$ = 7 meV measured along
the [010] cubic direction taken in the low temperature phase (400 K) and in the higher
temperature phase (650 K) are shown in Fig. 6. \ The solid lines are 
Gaussian fits that show the peak position or a ridge of scattering intensity
shifted to smaller $q$ as the temperature increases. \ These data, shown schematically in the inset of Fig.6, suggest that the ridge of scattering evolves into the expected TO phonon branch behavior at higher temperature. \ In the inset of Fig.6, the solid lines 

\begin{figure}[tb]
\epsfig{width=0.95\linewidth,figure=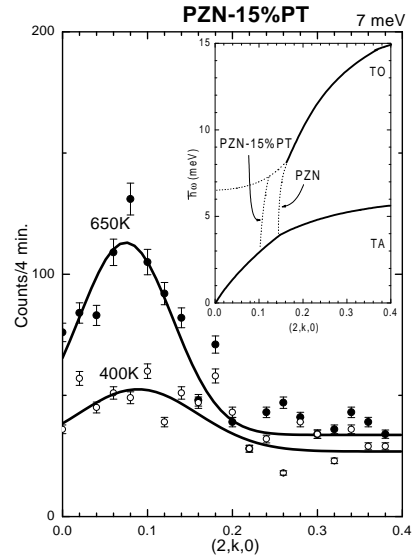}
\caption{Constant-$E$ scans at 7 meV (phonon creation) measured at 400 K and
650 K on a PZN-15\%PT crystal. \ Inset shows the dispersion curves for the TA and the
lowest-energy TO modes, showing the anomalous behavior of PZN-$x$PT. \ The data 
from Ref.\protect\onlinecite{Gehring_et_al_224109_2001} are also included and shown 
by two vertical dot lines.}
\label{fig6}
\end{figure}

\noindent represent dispersions of the transverse acoustic (TA) and the lowest-frequency TO phonon branches whereas the dashed lines show the presence of the steep ridge of scattering intensity at $q \sim 0.1$ recipocal lattice units (r.l.u.) for PZN-15\%PT and 0.14 r.l.u. for PZN at a temperature of about 500 K (see Ref.\onlinecite{Gehring_et_al_224109_2001}). It is interesting to note that the appearance of the steep ridge of scattering is shifted towards the zone-center i.e., (2,0.14,0), (2,0.13,0) and (2,0.1,0) for $x$=0, 8 and 15\%, respectively with increasing PT
concentrations. \ So far 15\%PT is the highest concentration that we have
studied, and it is possible that the higher concentrations might show a
softening of a zone-center TO phonon as observed in the conventional
ferroelectric PT\cite{Shirane_et_al_155_1970}, which would established the
end member of the waterfall. \ This indicates that the anomalous soft phonon
behavior is not simply a feature of compositions near the MPB but is
extended to large values of $x$. \ It must be related to the presence of
polar nanoregions rather than to the newly discovered phases. \ Therefore, the
microstructural characters of relaxor ferroelectrics are retained to some
degrees in the PZN-$x$PT solid solution systems up to a quite high
concentration of PT beyond the MPB.

\section{Discussion}

It now becomes clear that the presence of the new phase at the MPB in the
PZN-$x$PT phase diagram has orthorhombic rather than monoclinic symmetry. However, the orthorhombic cell observed here can be regarded as regared as a doubled 
monoclinic cell of the M$_{C}$ type (space group Pm) in the limit of $%
a=c,$ as discussed in Ref.\onlinecite{Noheda_et_al_3891_2001}. This is not the same type of monoclinic distortion M$_{A}$ that was observed in 
the PZT system, which has Cm symmetry. \cite{Noheda_et_al_PRB_14103_2001} \ 
The M$_{A}$ type has $b_{m}$ directed along the pseudo-cubic [110] direction, 
while in M$_{C}$ $b_{m}$ is directed along the pseudo-cubic [010].

In order to understand how the symmetry of the newly discovered phases facilitates 
the polarization rotation, it is useful to compare PZN-$x$PT with BaTiO$_{3}$. \ 
However, there is the question whether the new PZN-$x$PT
orthorhombic cell is fundamentally different from the old BaTiO$_{3}$
orthorhombic cell, both with the same space 

\begin{figure}
\epsfig{width=0.80\linewidth,figure=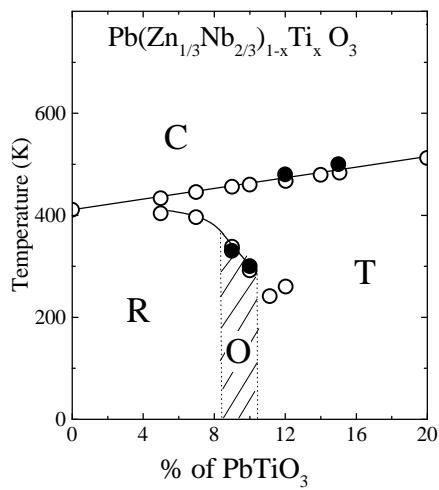}
\caption{Updated phase diagram of PZN-$x$PT around its MPB. The open
circles and solid lines represent the phase diagram by Kuwata et al. 
\protect\cite{Kuwata_et_al_579_1981}. \ The results of this work, as well as
those in Ref. \protect\onlinecite{Cox_et_al} ($x$=9$\%$)
are plotted as solid circles. \ The new orthorhombic phase (O) is
represented by the shaded-area.}
\label{fig7}
\end{figure}

\noindent group. \ The difference here seems to be in the very low energy barrier 
existing in PZN-$x$PT, between the O and the M$_{C}$ states, which allows the O 
polar axis [101] to rotate easily in the monoclinic plane.

The high-resolution synchrotron x-ray powder diffraction study of PZN-$x$PT (%
$10\%\leq x\leq 15\%$) as a function of temperature has allowed us to define 
the new orthorhombic
region of the phase diagram.\ We show that the low-temperature orthorhombic
structure of PZN-$x$PT is found for $x$= 10\% but not for higher
concentrations.\ With the results of this and earlier studies of PZN-$x$PT ($%
8\%\leq x\leq 15\%$), we have completed a revision of the PZN-$x$PT phase
diagram around its MPB as shown in Fig. 7. The new orthorhombic phase exists
in only a narrow concentration range ($8\%<x<11\%$) between the rhombohedral
and tetragonal phases with near vertical phase boundaries on both sides. \ These results, shown by the closed circles in Fig. 7, are in good agreement with those of Kuwata and
colleagues (shown by open circles) except for the two points indicating phase transitions at about 250 K for $x$ = 11\% and 12\%. \ One possible reason for this discrepancy might possibly be two-phase coexistence in their ceramic samples.

To be sure that the proposed phase diagram corresponds to the ground state
of these materials, the poled/unpoled problem deserves further attention.
Due to the intrinsic disorder existing in these relaxor ferroelectrics, the
diffraction peaks of the as-grown samples are very broad, and to solve the
true underlying structure by powder diffraction is an extremely difficult
task. \ In order to induce the ferroelectric ordered state the samples are
poled by an electric field. \ The extremely sharp peaks of the
ordered material allow us to unequivocally determine the structure. \
However, due to the small energy differences between the different phases
around the MPB, sometimes the poled and unpoled states of the samples do
not have the same symmetry or, alternatively the symmetry can depend on the
direction of the applied electric field, as in the case of the $x$=8\%
composition\cite{Noheda_et_al_3891_2001}. \ The as-grown sample is known to be
rhombohedral, but an irreversible phase transition can be induced by an
electric-field applied along the [001] direction, and the sample poled in
such a way becomes orthorhombic. However, the initial rhombohedral state can be
recovered by grinding the crystals below $\sim $30$\mu $m\cite
{Noheda_et_al_3891_2001}.

In a similar way, the $x$= 9\% composition showed very sharp and well-defined
diffraction peaks after poling, displaying a clear orthorhombic symmetry 
\cite{Cox_et_al}. In this case, however the poled and unpoled samples do not
have intrinsically different characteristics, as shown in Ref.\protect\onlinecite{Uesu_et_al._Cond-mat/00106552}, and the
main difference is the existence of a residual tetragonal phase in the
unpoled samples. \ Here it is worth mentioning that it is very difficult to
grow a perfectly homogeneous crystal and that certain compositional
variations are always present for the same nominal composition. \ In this
work we have shown that the poled $x$= 10\% crystal behaves much like 
the poled $x$=9\%\cite{Cox_et_al}, and that the poled $x$=11-15 \% samples are
tetragonal, as expected for the as-grown samples\cite{Kuwata_et_al_579_1981}%
. Therefore, we believe that the new PZN-$x$PT phase diagram contains an
orthorhombic intermediate phase for $8\%<x<11\%$ as shown in Fig. 7, where
the $x$ values stand for the nominal compositions.

For further characterization of PZN-$x$PT, we have subsequently performed a
neutron inelastic scattering on the PZN-15\%PT single crystal to study the
``waterfall'' feature. \ As was the case for PZN-8\%PT\cite
{Gehring_et_al_5216_2000} and later for PZN\cite{Gehring_et_al_224109_2001}
measured at the same temperature, an anomalous scattering ridge in
PZN-15\%PT is also found, but shifted closer to the zone center. \ It will
be interesting to study the crystals of higher values of $x$, and then to
trace the evolution of the TO modes as a function of $x$. \ Thus, the
conclusion of the present study is that the waterfall phenomenon is not
associated with the new low temperature phase near the MPB, but is a more
general feature of the PZN-$x$PT solid solution system, which retains some
degrees of the relaxor ferroelectric characters, to be associated with the
presence of the polar nanoregions.

During the preparation of this manuscript two different papers by Lu et al. 
\cite{Lu} and Xu et al \cite{Xu_et_al_020102_2001} have reported the optical
observation of orthorhombic and monoclinic domains, respectively, in the
related ferroelectric Pb(Mg$_{1/3}$Nb$_{2/3}$)O$_{3}$-$x$PbTiO$_{3}$ system
(PMN-$x$PT) around its MPB ($x\simeq 0.33-0.35\%$). A third paper by Ye et
al. \cite{Ye_et_al} reports the existence of a monoclinic phase of M$_{A}$
type in PMN-35\%PT, after the application of an electric field. Work is in
progress to clarify which type of low symmetry distortion occurs in the
phase diagram of PMN-xPT. \ After this paper was submitted, we became aware of a series of theoretical papers by Ishibashi and colleagues based on a Landau-Devonshire approach taken to fourth-order presenting a theory of the MPB in perovskite-type oxides, including calculations of dielectric susceptibilities and piezoelectric constants\cite{Ishibashi_etal}.

\acknowledgments
We wish to thank W. Chen for an excellent job in crystal growth. Research was carried out 
(in part) at the National Synchrotron Light Source, Brookhaven National Laboratory, which 
is supported by the US Department of Energy, Division of Material Science and Division 
of Chemical Science. We acknowledge the support of the NIST Center for Neutron Research, U.S.
Dept. of Commerce, in providing the neutron facilities used in this work. Financial
support by DOE under contracts No. DE-FG02-00ER45842 and No. DE-AC02-98CH10886 
and the ONR Grant \# N00014-99-1-0738 is also ackowledged.  %
%=================================Reference================================%

\end{document}